\documentclass[12pt]{article}

\usepackage{graphicx}
\begin{document}

\begin{center}

{\large\bf Morphology and Kinetics of Random Sequential Adsorption of Superballs: From Hexapods to Cubes}

\bigskip
Pooria Yousefi,$^1$ Hessam Malmir,$^{2,\dagger}$ and Muhammad Sahimi$^{3,\ddagger}$

{\it $^1$Faculty of Engineering, Science and Research Branch, Azad University, Tehran 14515-775, Iran\\
$^2$Department of Chemical and Environmental Engineering, Yale University, New Haven, Connecticut 06511, 
USA\\ $^3$Mork Family Department of Chemical Engineering and Materials Science, University of Southern 
California, Los Angeles, California 90089, USA}

\end{center}

Superballs represent a class of particles whose shapes are defined by ${|x|}^{2p}+{|y|}^{2p}+{|z|}^{2p}
\le R^{2p}$, with $p\in(0,\infty)$ being the {\it deformation parameter}. $0<p<0.5$ represents a family of 
hexapodlike (concave octahedrallike) particles, while for $0.5\leq p<1$ and $p>1$ one has, respectively, 
families of convex octahedrallike and cubelike particles, with $p=1,\;0.5$ and $\infty$ representing spheres, 
octahedra, and cubes. Colloidal zeolite suspensions, catalysis, and adsorption, as well as biomedical magnetic nanoparticles are but a few of the applications of packing of superballs. We introduce a universal method for simulating random sequential adsorption of superballs, which we refer to as {\it low-entropy} algorithm, in contrast with the conventional algorithm that represents a {\it high-entropy} method. The two algorithms yield, respectively, precise estimates of the jamming fraction $\phi_\infty(p)$ and $\nu(p)$, the exponent that characterizes the kinetics of adsorption at long times $t$, $\phi(\infty)-\phi(t)\sim t^{-\nu(p)}$. Precise estimates of $\phi_\infty(p)$ and $\nu(p)$ are obtained and shown to be in agreement, in some special limits, with the existing analytical and numerical results. 

\newpage

Random sequential adsorption (RSA) is an irreversible process for generating nonequilibrium packings of 
nonoverlapping particles, and is considered a very useful model to study and understand the structure of 
low-temperature phases of matter, as well as particle aggregation and jamming in a wide variety of 
applications, from granular media [1,2], to heterogeneous materials [3,4] and biological systems [5,6]. The 
RSA and its kinetics are also among important problems in statistical physics, which have been studied 
analytically and numerically for various particles and systems [1,5,7-17]. An important property of the RSA 
is the kinetics of the adsorption that typically approaches a very slow asymptotic saturation limit (jamming) 
in which no more particles can be added to the packing. 

In this Letter we focus on a special class of particles, the so-called {\it superballs}, whose possible
shapes include a variety of three-dimensional (3D) concave and convex particles. Colloidal zeolite suspensions
with applications in catalysis, adsorption, and separation [18,19], as well as packings of magnetic 
nanoparticles with biomedical applications [20-22] are but some of the better-known applications of the RSA of
superballs. Although optimal (lattice) packings and maximally randomly jammed (MRJ) systems of superballs have
been studied by Jiao {\it et al.} [23,24], packing of superballs by the RSA and its kinetics have not been 
investigated. The importance of modeling and analysis of the RSA of superballs is due to the fact that by 
tuning a shape parameter one obtains a wide variety of particle shapes, ranging from hexapod- to octahedral- 
and cubelike particles, as well as spherical ones. At the same time, determining the saturation coverage and 
kinetics of the RSA of various types of particles by a unified approach is a long-standing problem, which 
we address in this Letter by studying a large family of superballs. One of the main questions that we address 
is how changing of particles' shapes from concave to convex affects the maximum saturation coverage 
$\phi_\infty$ (sometimes called the jamming limit) of their packings. From practical view point, the 
adsorption rate of the particles is also an important property, which we study and compare with the existing 
conjectures on the kinetics of the RSA [1,2,5,7,8,10,13,15].

The shape of superballs is described by the following general equation,
\begin{equation}
\label{eq:superball}
{|x|}^{2p} + {|y|}^{2p} + {|z|}^{2p} \le R^{2p}\;,
\end{equation}
where $p\geq 0$ is the {\it deformation parameter} that indicates the extent to which the particle's shape 
deviates (deforms) from that of a sphere, the limit $p=1$, and $R$ is the superballs' radius. Depending on
$p$, a superball may possess two types of shape anisotropy, namely, cubelike and octahedrallike shapes. As 
$p$ increases from 1 to $\infty$, one obtains a family of convex superballs with cubelike shapes. The limit 
$p\to\infty$ represents a perfect cube. As $p$ decreases from 1 to 0.5, a family of convex superballs with 
octahedrallike shapes are obtained. In the limit $p=0.5$ the superballs represent regular octahedral particles.
For $p<0.5$, they still possess an octahedrallike shape, but similar to hexapods are concave, and approach a 
3D ``cross'' in the limit $p\to 0$.

The algorithm that we utilize for simulating the RSA of superballs is a generalization of the one that we
recently developed for cubic particles [25-27]. We begin with a large, empty box of volume $V$ in ${\bf R}^3$,
generate superballs with given deformation parameter $p$ and randomly-selected positions and orientations, and
place them sequentially in the simulation box. The deposition is subject to the nonoverlapping constraint, 
so that no newly inserted particle can overlap with any existing ones. The overlap occurrence depends, 
however, on the spatial coordinates and orientation of the superballs. Thus, (i) we first generate at random 
the three coordinates $x\in [0,x_{\rm max}]$, $y\in [0,y_{\rm max}]$ and $z\in [0,z_{\rm max}]$ of the 
superball's center of mass, where $(x_{\rm max},y_{\rm max},z_{\rm max})$ are the dimensions of the simulation
box, and (ii) the superball is then rotated using a quaternion [28], ${\bf q}=a+b{\bf i}+c{\bf j}+d{\bf k}$, 
which represents the orientations and rotations of 3D objects. A quaternion is simpler to compose than the 
Euler's angles that are used to describe the orientation of a rigid body; avoid losing one degree of freedom 
(the so-called gimbal lock problem [29]); are more compact than the rotation matrices and more stable 
numerically, and their use is more efficient. Henceforth, we need four independent random numbers $a,\;b,\;c$,
and $d$, uniformly sampled from a normal distribution with 0 mean and unit standard deviation in order to 
uniformly and randomly rotate the superballs. (iii) Periodic boundary condition is then imposed in all three 
directions. (iv) The surface points of the randomly positioned and rotated superball are then checked via Eq. 
(1) against the centers of the previously inserted superballs to see if any overlap occurs. If so, the 
superball is rejected and a new one is generated starting from step (i). Otherwise, the superball is accepted 
and the deposition process continues until the saturation or jamming limit of the system is reached.

The RSA rate is mainly limited by the volume exclusion from previously adsorbed particles. Its long-time 
kinetics is described by [1,10],
\begin{equation}
\label{eq:RSAkinetics}
\phi(\infty)-\phi(t)\sim t^{-\nu(p)}\;,
\end{equation}
in which $\phi(t)$ is the packing fraction at dimensionless time $t$:
\begin{equation}
\label{eq:packing_fraction}
\phi(t)=\frac{n(t)V_{sb}}{V}\;,
\end{equation}
and $\phi(\infty)$ denotes the RSA maximum saturated packing fraction, with $V$ being the simulation box's 
volume. $n(t)$ is the number of superballs generated up to time $t$, and $V_{sb}$ refers to the volume of the 
superballs [23].

Since the positions of superballs in an RSA packing are equiprobable throughout the simulation box, random 
sampling of a superball's position follows a uniform probability density function (PDF). Thus, the 
predictability of the existence of an empty space for inserting a superball, which is a random variable 
$X$ with a uniform distribution is controlled by the interval in which the PDF is nonzero, with the simulation
box size being $L=\max[x_{\rm max},y_{\rm max},z_{\rm max}]$. For $L\to 0$, the PDF becomes a delta function 
and the predictability is maximal, i.e., the uncertainty is minimum. In this limit, $X$ takes on the value at 
which the delta function is nonzero. For $L\to\infty$, however, the predictability of the state of $X$ is 
minimal, i.e., the uncertainty is maximum, and the same is true for all the possible states. Thus, one 
requires a measure of the uncertainty for the state of the random variable $X$ [30]. Let $X$ be a discrete 
random variable with possible values $\{x_1,\cdots,x_m\}$ and probabilities $P_i=P(X=x_i)$. The entropy $S$, 
a measure of uncertainty, is defined as the expected value of the information gained from observing $X$ [31]:
\begin{equation}
S=-\sum_i P_i\log P_i={\cal E}[-\log P(X)]\;,    
\end{equation}
where ${\cal E}$ is the expected value operator. Thus, $S$ depends on the probability distribution of 
$P_1,\cdots,P_m$, but not on $x_1,\cdots,x_m$. Applying the definition to the RSA in a simulation box of 
length $L$ yields
\begin{equation}
S= -\int_0^L\frac{1}{L}\ln\left(\frac{1}{L}\right)dx=\ln L\;,    
\end{equation}
implying that by shrinking the size $L$ of the  sampling domain, the uncertainty $S$ of finding empty spaces 
in the RSA also decreases. But, the questions are, how much can one possibly limit the domain of the sampling,
and how does it affect the maximum RSA packing fraction $\phi_\infty$ and its kinetics, which should be 
independent of the simulation box's dimensions?

To address the questions, we must consider an approach that satisfies the isotropy of the packings and does 
not alter the RSA constraints. To do so, we propose to first sample the entire simulation box in order to 
randomly distribute the superballs, and achieve a state in which the number of iterations to find an empty 
space for inserting a superball becomes very large. We refer to this step as {\it phase I}. We then divide 
the simulation box into uniform and equal grid cells, with their size selected such that they can accommodate 
a few superballs, at least 3 or 4. We then sweep the cells one by one, referred to as {\it phase II}, to more 
accurately identify any possible empty space that can accomodate new superballs. By decreasing the domain of 
the sampling in phase II, the predictability of finding empty space for particle insertion increases, ensuring
that the true saturation is reached. 

We refer to the combination of the two phases as the {\it low-entropy} RSA, which yields precise estimates
of the maximum saturation coverage, $\phi_\infty$. This approach cannot, however, capture accurately and 
efficiently the kinetics of the RSA because phase II that explores the cells sequentially is a very slow 
process. Thus, it makes reaching the true asymptotic kinetics of the RSA difficult. Hence, we still need to 
use the conventional RSA, which we refer to as a {\it high-entropy} RSA process, in which phase II is not 
considered, or has a negligible effect in the simulations and, therefore, almost all the superballs are 
generated and inserted by phase I. The use of low- and high-entropy terminology is motivated by Eq. (5),
according to which by limiting the size $L$ of the domain of sampling, entropy $S$ is kept very low, and the absence of this effect in the high-entropy case.

\begin{figure}[htbp]
\centering
\includegraphics[width=\linewidth]{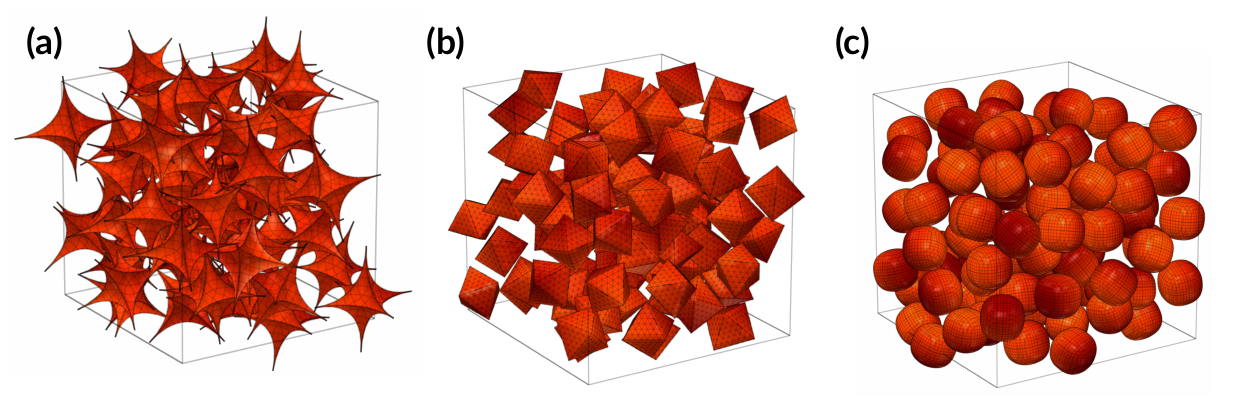}
\vspace{-20pt}
\caption{Packings of (a) concave superballs with $p=0.25$; (b) octahedral particles with $p=0.5$, and (c) 
cubelike particles with $p=1.25$. The packings are subject to periodic boundary conditions.}
\end{figure}

Figure 1 shows three RSA packings. The densest packing in Fig. 1 is that of cubelike particles with $p=5/4$
whose shape is still close to spherical particles, $p=1$. Using the algorithms and setting $R=1$ for the 
superballs, we computed precise estimates for $\phi_\infty$ and the kinetics of adsorption for ten types of 
superballs as a function of the deformation parameter $p$, ranging from $p=0.25$ (concave octahedrallike 
particles), to 0.5 (octahedra), 0.75 (convex octahedrallike particles), and 1 (spheres), as well as 6 values 
of $p$ for the cubelike superballs, using the low-entropy RSA algorithm. The results are shown in Fig. 2. The 
maximum value of $\phi_\infty(p)$ corresponds to spherical particles, for which we obtained, $\phi_\infty 
\approx 0.384457\pm 0.003991$, which is in agreement with the previous estimates, $\phi_\infty\approx 0.38278$
[12], 0.3841307 [13], 0.38 [32], and 0.382 [33]. As $p\to 0$, $\phi_\infty(p)$ decreases, with the RSA packing
of hexapodlike superballs with $p=0.25$ having the lowest $\phi_\infty$ that we computed, $\phi_\infty \approx
0.151749\pm 0.001553$. Beyond spheres we have cubelike particles with $p>1$ for which $\phi_\infty$ also 
decreases, and for $p>10$ reaches the saturation limit, $\phi_\infty\approx 0.333=1/3$ for the cubic particles
[34]. This is presumably the most accurate estimate of $\phi_\infty$ for the RSA packing of cubes, since 
preventing overlaps between the particles is based on Eq. (1) as $p\to\infty$, and not based on approximations
in terms of edge-edge, edge-face, and corner-face intersections [16].

The plot of $\phi_\infty$ versus $p$ shown in Fig. 2 has a shape distinctly different from that of the lattice
and MRJ packings of superballs [23,24], for which $\phi_\infty$ has its lowest value for the spherical 
particles, $p=1$. This demonstrates that the RSA is a process completely different from those that generate 
dense equilibrated configurations of superballs. The list of estimates of $\phi_\infty(p)$ along with their 
standard errors, computed by using the low-entropy RSA, is presented in Table 1. The fitted curve in Fig. 2 
may be approximated by 
\begin{equation}
\phi_\infty(p)=C_1\exp(-C_2p)\cos(C_3 p-C_4)+C_5\;,
\end{equation}
with $C_1$, $C_2$, ..., $C_5$ being, respectively, $\approx 1.661$, 2.089, 0.567, 1.901, and 0.336. 

\pagebreak

\begin{center}
Table 1. Maximum packing fraction $\phi_\infty$ of superballs versus the deformation parameters $p$.

\bigskip

\begin{tabular}{cc} 
\hline
$p$ & $\phi_\infty(p)$ \\
\hline
1/4 & $0.151749\pm 0.001553$ \\
1/2 & $0.307333\pm 0.002553$ \\
3/4 & $0.370720\pm 0.005747$ \\
1   & $0.384457\pm 0.003991$ \\
5/4 & $0.378758\pm 0.002304$ \\
2   & $0.355377\pm 0.005804$ \\
6   & $0.340227\pm 0.002650$ \\
10  & $0.335687\pm 0.004656$ \\
50  & $0.333634\pm 0.006789$ \\
100 & $0.333832\pm 0.003552$ \\
\hline
\end{tabular}
\end{center}

\begin{figure}[htbp]
\centering
\includegraphics[width=12cm]{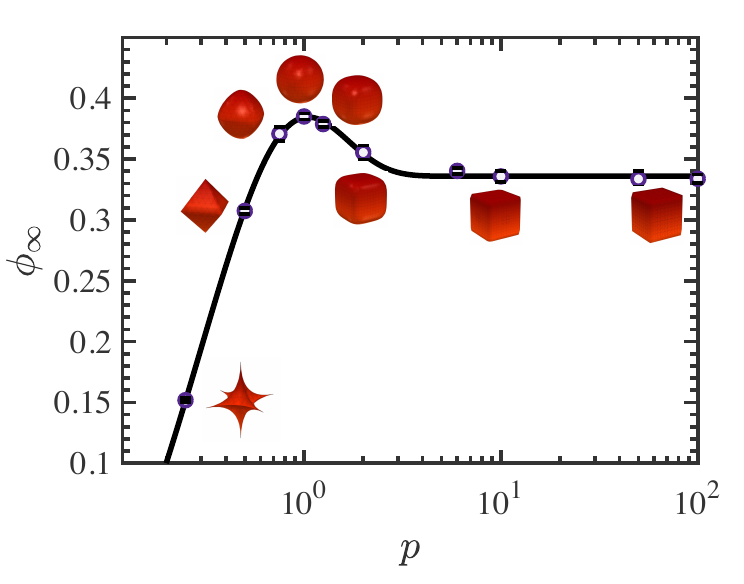}
\vspace{-15pt}
\caption{Maximum RSA packing fractions versus the deformation parameter $p$. For $p\geq 5$, the shape of 
particles approaches that of cubes with their maximum RSA packing fraction being $\sim 0.333$.}
\end{figure}

\bigskip

To obtain accurate results for the kinetics of the RSA packing, Eq. (2), we use the high-entropy RSA 
algorithm. Figure 3 shows the dependence of $\phi(2t)-\phi(t)$ on the dimensionless time $t$ defined by, $t=
n_iV_{sb}/V$ in which $n_i$ denotes the number of the RSA iterations, i.e. the number of successive
addition of the particles. Since it is not practical to carry out the simulations for too long, we derive the
asymptotic behavior by analyzing $\log[\phi(2t)-\phi(t)]$ that exhibits the same scaling as $\log[\phi(\infty)
-\phi(t)]$, when plotted against $\log(t)$ [15]. As illustrated, the slope of $\log[\phi(2t)-\phi(t)]$, which 
corresponds to $-\nu(p)$ in Eq. (2), has its maximum and minimum values at, respectively, $p=1$ (spheres) and 
$p=0.25$ (hexapodlike superballs). For spherical particles, one has, $\nu\approx 0.33$, which was previously 
predicted [5,7,8,12,30]. For cubes and octahedra, however, $\nu\approx 0.23$ and 0.18, respectively. The 
plot of $\log[\phi(\infty)-\phi(t)]$ with respect to $\log(t)$ decays slowly for concave octahedrallike 
particles. For $p=0.25$ we obtain $\nu\approx 0.07$. In this limit, the relation $\phi(\infty)-\phi(t)\sim
\log(t)$ also quantifies the asymptotic behavior extremely accurately, which is not surprising as $\nu$ 
is very small.

\begin{figure}[htbp]
\centering
\includegraphics[width=12cm]{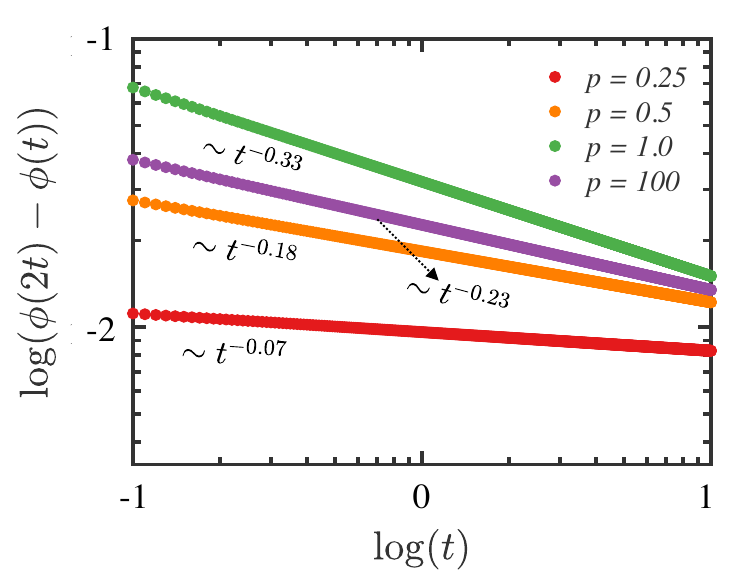}
\vspace{-15pt}
\caption{The kinetics of the RSA for various deformation parameters $p$. The asymptotic behavior of 
$\log[\phi(2t)-\phi(t)]$ that exhibits the same scaling as $\log[\phi(\infty)-\phi(t)]$ when plotted against 
$\log(t)$ [Eq. (2)]. Purple, green, and orange show, respectively cubes, spheres, and octahedra, while red 
represents hexapodlike particles with $p=0.25$.}
\end{figure}

To further characterize the structure of the RSA packings, we computed the pair correlation function $g_2(r)$
at $\phi_\infty(p)$. Figure 4 presents the results. The peak of $g_2(r)$ occurs at $r=D$ for the spherical 
particles, where $D=2R$ is the superballs' diameter, and moves from left to right with increasing $p$. Since 
for superballs with $p<1$ the diameter $D$ equals the diameter of their circumscribed sphere, Fig. 4 indicates
that for $p=0.25$ and $p=0.5$ (octahedron) and $r<D$ we still have some neighboring superballs around the 
reference one, but the peak of $g_2(r)$ occurs at smaller $r$ for smaller $p$, and has lower values for such 
values of $p$. On the other hand, for superballs with $p>1$ the diameter $D$ is the diameter of their 
inscribed sphere, which is why the peak of $g_2(r)$ in Fig. 4 for $p=2.0$ (cubelike particles) occurs at 
$r>D$, although the value of the peak is still lower than that of spheres ($p=1$). Furthermore, the limiting 
values for $g_2(r)$ of spheres are similar to those of Refs. [7,8,12,13], namely, $\lim_{r\to D^+}g_2(r)\sim 
-\ln{(r/D-1)}$, which supports the conjecture on the logarithmic singularity of $g_2(r)$ at $r=D$ for 
spherical particles.

\begin{figure}[htbp]
\vspace{2pt}
\centering
\includegraphics[width=12cm]{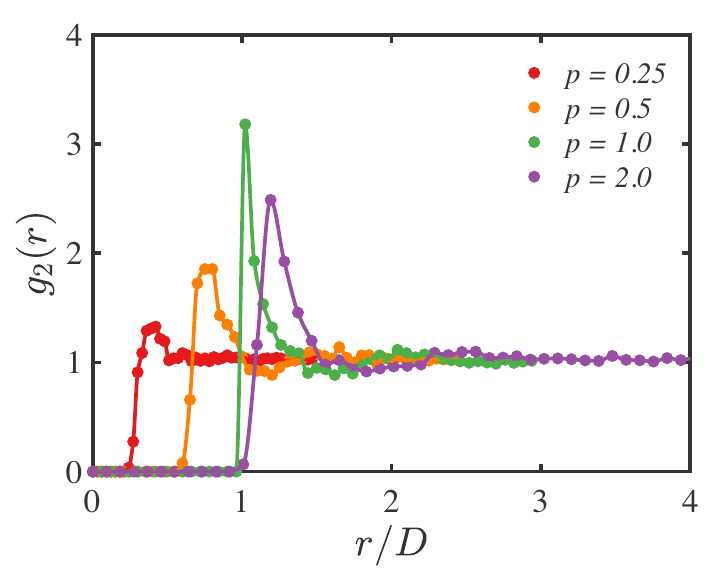}
\vspace{-15pt}
\caption{The pair correlation function $g_2(r)$ for the RSA packing of superballs at $\phi_\infty$. Purple, 
green, and orange show, respectively, cubelike particles ($p=2$), spheres ($p=1$), and octahedra ($p=0.5$),
while red shows the results for hexapodlike particles ($p=0.25$). The peak of $g_2(r)$ moves from left to 
right as the deformation parameter $p$ increases. The pair correlation function for the RSA of spherical 
particles follows, $\lim_{r\to D^+}g_2(r)\sim -\ln{(r/D-1)}$.}
\end{figure}

Summarizing, developing a new simulation algorithm, we presented the results of a comprehensive study of the 
RSA of superballs. The maximum packing fraction and the kinetics of the adsorption were studied and the 
relevant quantities were estimated. We proposed a universal approach, namely, the low-entropy process that 
leads to the precise estimates of the jamming. The highest saturated packing fraction among superballs belongs
to spherical particles ($p=1$). Both the long-time kinetics and the pair-correlation function $g_2(r)$ of the 
RSA of superballs at $p=1$ support the previous conjectures on the RSA of spherical particles. 

Work at USC was supported in part by the Petroleum Research Fund, administered by the American Chemical 
Society.

\bigskip

\bigskip

\noindent $^\dagger$hessam.malmir@yale.edu

\noindent $^\ddagger$moe@usc.edu

\newcounter{bean}
\begin{list}%
{[\arabic{bean}]}{\usecounter{bean}\setlength{\rightmargin}{\leftmargin}}

\item J.W. Evans, Random and cooperative sequential adsorption, Rev. Mod. Phys. {\bf 65}, 1281 (1993).

\item S. Torquato and F.H. Stillinger, Jammed hard-particle packings: From Kepler to Bernal and beyond, Rev. 
Mod. Phys. {\bf 82}, 2633 (2010).

\item S. Torquato, {\it Random Heterogeneous Materials} (Springer, New York, 2002).

\item M. Sahimi, {\it Heterogeneous Materials I} (Springer, New York, 2003).

\item J. Feder, Random sequential adsorption, J. Theor. Biol. {\bf 87}, 237 (1980).

\item J. Talbot, G. Tarjus, P.V. Tassel, and P. Viot, From car parking to protein adsorption: an overview of 
sequential adsorption processes, Colloids Surfaces A {\bf 165}, 287 (2000).

\item Y. Pomeau, Some asymptotic estimates in the random parking problem, J. Phys. A {\bf 13}, L193 (1980).

\item R.H. Swendsen, Dynamics of random sequential adsorption, Phys. Rev. A {\bf 24}, 504 (1981).

\item E.L. Hinrichsen, J. Feder, and T. J{\o}ssang, Random packing of disks in two dimensions, Phys. Rev. A 
{\bf 41}, 4199 (1990).

\item P. Viot and G. Tarjus, Random sequential addition of unoriented squares: Breakdown of Swendsen's 
conjecture, Europhys. Lett. {\bf 13}, 295 (1990).

\item J.-S. Wang and R.B. Pandey, Kinetics and jamming coverage in a random sequential adsorption of polymer 
chains, Phys. Rev. Lett. {\bf 77}, 1773 (1996).

\item S. Torquato, O.U. Uche, and F.H. Stillinger, Random sequential addition of hard spheres in high 
Euclidean dimensions, Phys. Rev. E {\bf 74}, 061308 (2006).

\item G. Zhang and S. Torquato, Precise algorithm to generate random sequential addition of hard hyperspheres 
at saturation, Phys. Rev. E {\bf 88}, 053312 (2013).

\item Y.Y. Tarasevich, V.V. Laptev, N.V. Vygornitskii, and N.I. Lebovka, Impact of defects on percolation 
in random sequential adsorption of linear $k$-mers on square lattices, Phys. Rev. E {\bf 91}, 012109 (2015).

\item A. Baule, Shape universality classes in the random sequential adsorption of nonspherical particles,
Phys. Rev. Lett. {\bf 119}, 028003 (2017).

\item M. Ci\'esla and P. Kubala, Random sequential adsorption of cubes, J. Chem. Phys. {\bf 148}, 024501 
(2018).

\item G. Zhang, Precise algorithm to generate random sequential adsorption of hard polygons at saturation,
Phys. Rev. E {\bf 97}, 043311 (2018).

\item P. Sharma, J.-S. Song, M.H. Han, and C.-H. Cho, GIS-NaP1 zeolite microspheres as potential water 
adsorption material: Influence of initial silica concentration on adsorptive and physical/topological 
properties, Sci. Rep. {\bf 6}, 22734 EP (2016).

\item S. Li, J. Li, M. Dong, S. Fan, T. Zhao, J. Wang, and W. Fan, Strategies to control zeolite particle 
morphology, Chem. Soc. Rev. {\bf 48}, 885 (2019).

\item G. Salazar-Alvarez, J. Qin, V. Sepel\'ak, I. Bergmann, M. Vasilakaki, K.N. Trohidou, J.D. Ardisson, 
W.A.A. Macedo, M. Mikhaylova, M. Muhammed, M.D. Bar\'o, and J. Nogu\'es, Cubic versus spherical magnetic 
nanoparticles: The role of surface anisotropy, J. Amer. Chem. Soc. {\bf 130}, 13234 (2008).

\item J. Zhang, J. Du, Y. Qian, Q. Yin, and D. Zhang, Shape-controlled synthesis and their magnetic properties
of hexapod-like, flake-like and chain-like carbon-encapsulated Fe$_3$O$_4$ core/shell composites, Mater. Sci. 
Eng. B {\bf 170}, 51 (2010).

\item J.G. Donaldson, P. Linse, and S.S. Kantorovich, How cube-like must magnetic nanoparticles be to modify
their self-assembly? Nanoscale {\bf 9}, 6448 (2017).

\item Y. Jiao, F.H. Stillinger, and S. Torquato, Optimal packings of superballs, Phys. Rev. E {\bf 79}, 
041309 (2009).

\item Y. Jiao, F.H. Stillinger, and S. Torquato, Distinctive features arising in maximally random jammed 
packings of superballs, Phys. Rev. E {\bf 81}, 041304 (2010).

\item H. Malmir, M. Sahimi, and M.R. Rahimi Tabar, Microstructural characterization of random packings of 
cubic particles, Sci. Rep. {\bf 6}, 35024 (2016).

\item H. Malmir, M. Sahimi, and M.R. Rahimi Tabar, Packing of nonoverlapping cubic particles: Computational 
algorithms and microstructural characteristics, Phys. Rev. E {\bf 94}, 062901 (2016).

\item H. Malmir, M. Sahimi, and M.R. Rahimi Tabar, Statistical characterization of microstructure of packings
of polydisperse hard cubes, Phys. Rev. E {\bf 95}, 052902 (2017).

\item J. Conway and D. Smith, {\it On Quaternions and Octonions} (Taylor \& Francis, London, 2003).

\item E.G. Hemingway and O.M. O'Reilly, Perspectives on Euler angle singularities, gimbal lock, and the 
orthogonality of applied forces and applied moments, Multibody Syst. Dyna. {\bf 44}, 31 (2018).

\item S. Heinz, {\it Mathematical Modeling} (Springer, Berlin, 2011).

\item M. Kelbert and Y. Suhov, {\it Information Theory and Coding by Example} (Cambridge University Press, 
London, 2013).

\item D.W. Cooper, Random-sequential-packing simulations in three dimensions for spheres, Phys. Rev. A 
{\bf 38}, 522 (1988).

\item J. Talbot, P. Schaaf, and G. Tarjus, Random sequential addition of hard spheres, Mol. Phys. {\bf 72}, 
1397 (1991).

\item In Refs. [25,26] the estimate $\phi_\infty\approx 0.57$ was reported for the RSA packings of cubes. The 
estimate was, however, erroneous because upong repeating the calculations with the present method it was
discovered that even very small overlaps between the particles, as small as 1-2\%, change $\phi_\infty$ very
significantly. The estimate we present in this Letter agrees with that of Ref. [16].

\end{list}%

\end{document}